\documentclass[tighten, twocolumn]{aastex63}
\usepackage{xcolor, fontawesome, microtype, amsmath}
\definecolor{twitterblue}{RGB}{64,153,255}
\newcommand{\twitter}[1]{\href{https://twitter.com/#1}{\textcolor{twitterblue}{\faTwitter}\,\tt \textcolor{twitterblue}{@#1}}}
\shorttitle{Searching for a message in the angular power spectrum of the cosmic microwave background}
\shortauthors{Michael Hippke}
\begin{document}
\title{Searching for a message in the angular power spectrum\\of the cosmic microwave background}
\author[0000-0002-0794-6339]{Michael Hippke}
\affiliation{Sonneberg Observatory, Sternwartestr. 32, 96515 Sonneberg, Germany \twitter{hippke}}
\affiliation{Visiting Scholar, Breakthrough Listen Group, Berkeley SETI Research Center, Astronomy Department, UC Berkeley}
\email{michael@hippke.org}

\begin{abstract}
The Creator of the universe could place a message on the most cosmic of all billboards, the Cosmic Microwave Background (CMB) sky. It was proposed by \citet{2006MPLA...21.1495H} to search for such a message in the CMB angular power spectrum. I process the temperature measurements taken by the Planck and WMAP satellites and extract the binary bit-stream. I estimate the information content of a potential message in the stream as about $1{,}000$\,bits. The universality of the message may be limited by the observer-dependent location in space and the finite observation time of order 100\,bn years after the big bang. I find no meaningful message in the actual bit-stream, but include it at the end of the manuscript for the interested reader to scrutinize.
\end{abstract}


\section{Introduction}
After the Big Bang, our universe was hot and dense, and photons where glued to matter. It cooled while expanding, and when its temperature dropped below $3{,}000$\,K, the photons were released. Today, we see these photons as the cosmic microwave background (CMB), a 2.7\,K electromagnetic radiation. As the universe was not perfectly uniform, density perturbations lead to a temperature anisotropy (directional dependency) of about $30\,\mu$K. These all-sky fluctuations can be converted to a power spectrum. Its angular wavenumber is referred to as the multipole $\ell$. The angular scale is approximately $\theta=180^{\circ}/\ell$, so that $\ell=2$ corresponds to about $90^{\circ}$. The angular power spectrum shows prominent ``acoustic oscillations'' which correspond to density perturbations of particles in the early universe. These features are used to constrain the cosmological parameters such as the matter and baryon density and the Hubble constant $H_0$.

A few unusual temperature variations in the CMB sky are described as anomalies, the most prominent being the ``Cold Spot'' \citep{2005MNRAS.356...29C}. There is a fine line between the (most likely) physical explanations for these features, the claim of concentric circles in the noise \citep{2010arXiv1011.3706G}, and April Fool's jokes of emoticons painted into the sky \citep{2011arXiv1103.6262Z,2016arXiv160309703F}.

In contrast to these not always serious attempts of finding anomalies in the CMB sky map, an earnest investigation into the CMB power spectrum was proposed by \citet{2006MPLA...21.1495H}. Their assumptions were, first, that some superior Being created the universe. Second, that the Creator actually wanted to notify us that the universe was intentionally created. Then, the question is: How would they send a message? The CMB is the obvious choice, because it is the largest billboard in the sky, and is visible to all technological civilizations. \citet{2006MPLA...21.1495H} continue to argue that a message in the CMB would be identical to all observers across space and time, and that the information content can be reasonably large (thousands of bits). In this paper, I discuss these claims using theoretical arguments.  For the first time, I examine power spectrum data (from the Planck and WMAP satellites) to actually search for such a message.

\section{CMB evolution over cosmic time}
As the universe cools, the CMB sky map and its resulting power spectrum are not static over time. The CMB temperature has already dropped from $3{,}000$\,K to 2.7\,K over 13\,bn years, and will continue to decrease with the cosmological scale factor $a$ as $1/a$. Then, the anisotropy power drops as $1/a^2$ which results in a shift of power spectrum features (the peaks and troughs) uniformly. These shift to smaller angular scales (larger multipoles) as time progresses. We could measure the Hubble constant $H_0$ from the cooling of the CMB monopole given enough time or precise instruments \citep{2020ApJ...893...18A}. After infinite time, the sphere of the last scattering surface approaches a maximum comoving radius, so that its features are of a (finite) minimum size. Today, the first acoustic peak is located at $\ell_0=221$ and will be observed at $\ell_f=290$ after infinite time . The stretch factor between now and the infinite future, for all features in the CMB, is $\ell_f=1.31 \ell_0$ \citep{2007PhRvD..76l3010Z,2008PhRvD..77d3505M}.

When the CMB temperature decreases, its wavelength increases. The peak in the 2.7\,K blackbody spectrum is today at a wavelength of a few mm, and will increase to 1\,m in 100\,Gyr. Its amplitude will then have decreased (been redshifted) by 12 orders of magnitude, making it difficult to detect \citep{2007GReGr..39.1545K}. This epoch will still be well within the Stelliferous Era, where stars shine and intelligent life (as we know it) can exist. Yet, these future inhabitants of the universe may be unable to detect the CMB. At the latest, the CMB will become undetectable when the wavelength of its photons are longer than the horizon size ($\approx 3{,}000$\,Mpc) in $\approx10^{40}\,$yrs \citep{1997RvMP...69..337A}. 

In the following, we will neglect the long-term cooling which makes detection more difficult, but will consider the evolution in the $C_{\ell}$ as these are relevant over short cosmological times (Gyr).

\section{The CMB for observers in different locations}
All technical civilizations in all galaxies in the universe can observe the CMB. But do they see the same CMB? It is clear that the sky map of 3D thermal fluctuation patterns is different for observers in different locations, since their past light cones sample different portions of the surface of last scattering. But perhaps observers separated by cosmological distances still see the same set of $C_{\ell}$ in the power spectrum? This would be useful for the purpose of a universal message (although one could argue that we live in a simulation \citep{Bostrom2003}, and the message is just for us).

\citet{2006MPLA...21.1495H} argue that the $C_{\ell}$ are the same for distant observers. This may be incorrect. For example, the \citet{1967ApJ...147...73S} effect (photons from the CMB are gravitationally redshifted) dominates the power spectrum for scales $>2^{\circ}$ ($\ell<90$). Its power depends on the location (observer) dependent surface of last scattering, so that different observers (separated by cosmological distances) see different values for their $C_{\ell}$ \citep{2005physics..11135S}. It is difficult to estimate the strength of the observer-dependent effect for smaller scale perturbations (higher $\ell$). However, we can discuss two scenarios.

First, we can consider that the Creator placed a message in the CMB which is \textit{different} for observers in different locations. As a naive example, Earthlings could find $\pi$ in the sky, while the inhabitants of a distant galaxy would see Euler's $e$ in the sky.

Alternatively, we can examine the idea that observers across all locations see the \textit{same} $C_{\ell}$. While their 3D CMB sky will be very different, there are infinitely many variations of the sky map which all collapse into the \textit{same} power spectrum. As an analogy, there are infinitely many functions which have the same integral (or derivative).

The problem with this second approach is that it causes an isotropy violation, i.e. that expectation values of the CMB temperature fluctuations are assumed to be preserved under rotations of the sky \citep{2003ApJ...597L...5H}. Does our CMB sky actually violate isotropy? Apparently yes, at the 1\,\% significance level in WMAP \citep{2012A&A...544A.121S} and Planck data \citep{2016CQGra..33r4001S,2018MNRAS.475.4357R}. It might be due to random noise \citep{2019JCAP...08..007S}, uncorrected galactic foregrounds, or unknown instrumental systematics. In the future, new polarization data could resolve the question (see section 2.7 in \citet{2019arXiv190602552P} and \citet{2016PhRvL.116v1301M}). Isotropy violations, if truly existent, could indicate either unexplored physics \citep[e.g. strings,][]{ 2017PhRvD..96b3512J} in the early universe, or the presence of a message in the CMB.

\begin{figure}
\includegraphics[width=\linewidth]{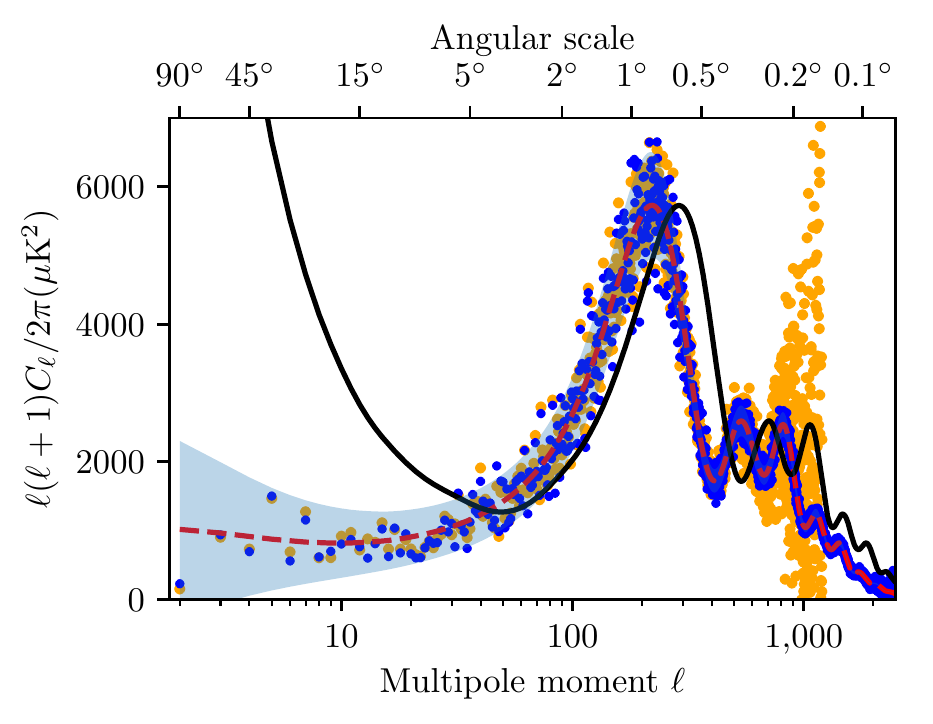}
\caption{\label{fig:cmb1}Angular power spectrum of the CMB. Data from \textcolor{blue}{Planck} and \textcolor{orange}{WMAP} with best-fitting 6-parameter cold dark matter ($\Lambda$CDM) \textcolor{red}{model}. In the infinite future, the spectrum will appear as the black line. \textcolor{cyan}{Cosmic variance} limits the significance.}
\end{figure}

\section{Sample variation and cosmic variance}
Two effects limit the accuracy of data obtainable from experiments. The first is the \textit{sample variation}: Due to the galactic foregrounds, only part of the CMB can be observed -- for Planck, this was 65\,\%. The observed sample is only part of the parent population and introduces an uncertainty into the number estimates. This uncertainty could be reduced by observations made from outside of our galaxy.

The second effect is the \textit{cosmic variance}: We have only one sky to measure. This sampling uncertainty is due to the fact that each $C_{\ell}$ is $\chi^2$ distributed with $2 \ell + 1$ degrees of freedom. For example, for $C_{\ell}=2$ there are only 5 coefficients, and a variance estimate
from 5 numbers is statistically uncertain. The unavoidable estimation error for every $C_{\ell}$ is $\Delta C_{\ell}=\sqrt{2/(2\ell +1)}$ (shaded area in Figure~\ref{fig:cmb1}). The cosmic variance is smaller at small angular scales, because there are more regions on the sky that can be considered independent samples of the parent distribution. Observers from different locations in the universe would see the surface of last scattering differently, and could provide independent observations, reducing cosmic variance \citep{2019PhRvD.100b3518L}. Sensitivity of Planck data was higher than the limit from cosmic variance up to $C_{\ell}\sim1{,}600$ \citep[p. 45 in][]{2018arXiv180706205P}.

\begin{figure}
\includegraphics[width=\linewidth]{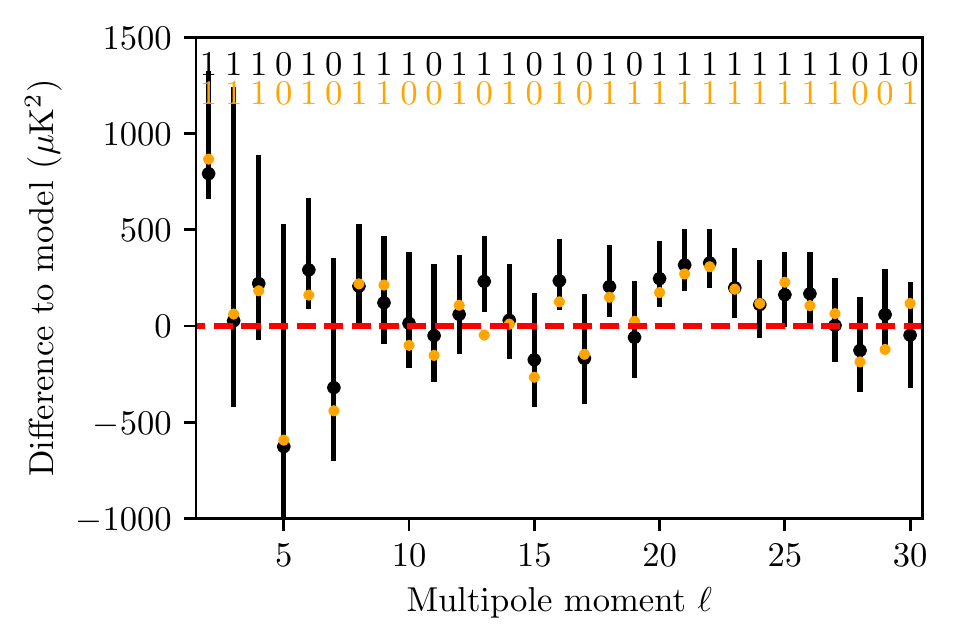}
\caption{\label{fig:cmb2}CMB data from \textcolor{blue}{Planck} and \textcolor{orange}{WMAP} divided by the best-fitting 6-parameter model \textcolor{red}{model}. Values above the mean are denoted as ``1'', below as ``0''. Error bars include cosmic variance.}
\end{figure}

\section{Message encoding and information content}
The presence of sample variation and cosmic variance limit the number of possible encodings per $C_{\ell}$, and thus the information content of a CMB message. This is a plausible limitation in the scenario where all inhabitants of the universe see a different CMB sky but find that it collapses to a very similar power spectrum (at the cost of isotropy violation). In this case, the power spectrum might not be perfectly identical, just similar enough to contain the same message. As an example, we might observe $C_2=700\pm5\,\mu$K$^2$, while a distant observer would see $C_2=800\pm5\,\mu$K$^2$. Both values could then encode the same letter ``A''.

It was suggested by \citet{2006MPLA...21.1495H} that a ternary encoding is most natural: values above, below and equal to the model. Three states are used in group theory in its application to particle physics. This raises the question of the steps involved: How much deviation is required for a value to be above the expectation? In Planck data, there is no indication of \textit{steps} in units of $\mu$K$^2$ (compare Figure~\ref{fig:cmb2}) which could be indicative of an encoding of $n$ bits per residual (e.g., $+800\,\mu$K$^2$, $+700\,\mu$K$^2$ etc.). I find the distribution of residuals to be approximately Gaussian with no indication of multi-modality. As Planck data are already limited by cosmic variance, a possible ternary coding should be visible. As that is not the case, it seems more natural to assume a binary encoding where (arbitrarily) a \texttt{1} is assigned to positive and \texttt{0} to negative residuals.

How much data can be encoded in the power spectrum assuming a binary message? A first limit comes from the fact that subsequent $C_{\ell}$ are in theory causally correlated. In practice, the strength of the effect is difficult to quantify, and a prominent lack of correlation is noted by the \citet{2019arXiv190602552P}. The smoothing scale has been estimated as $\Delta \ell = 10$ \citep{2006MPLA...21.1495H}. In addition, the number of  $C_{\ell}$ available for encoding is limited to $\ell \lesssim 2{,}000$. Smaller spatial variations are greatly smoothed and suppressed due to the finite thickness of the last scattering surface. With these limitations, the information content is 

\begin{equation}
M = \frac{\ell_{\rm max}}{2 \Delta \ell} \ln \left( \frac{\ell_{\rm max}}{2} \right)\,\,{\rm bits}.
\end{equation}

Taking $\ell_{\rm max}=2{,}000$ and $\Delta \ell = 10$ we get $M=660\,$bits. A shorter smoothing scale and a higher $\ell_{\rm max}$ increase this number. For example, there is still some variance visible at $C_{\ell}=2{,}500$, and one could argue for $\Delta \ell=5$; a combination which gives $M=1{,}700$. It is clear that $M \approx 1{,}000\,$bits within a factor of a few.

\section{Message}
With all the disclaimers out of the way, we can proceed to pull the bit-stream from the CMB data and test whether it contains a (meaningful) message. Temperature fluctuations in the microwave sky have been detected by the COBE mission \citep{1992ApJ...396L...1S} and examined with increasing accuracy by WMAP \citep{2013ApJS..208...20B,2013ApJS..208...19H} and Planck \citep{2018arXiv180706205P,2019arXiv190712875P}. Starting with the Planck and WMAP data as shown in Figure~\ref{fig:cmb1}, we subtract the model from both datasets. The residuals are shown in Figure~\ref{fig:cmb2}. 
The resulting binary message from Planck and WMAP is as follows. Values found identical in both missions are printed in black, deviating values in \textcolor{red}{red} (choosing Planck) . Errors for $\ell<1{,}600$ are dominated by cosmic variance. I estimate black values to be \textit{true} with a probability of $\sim90\,$\% on average, those in red with $\sim60\,$\%. Out of the first 100 bits, 83 are identical between Planck and WMAP; 409 from the first 500 (82\,\%); and 703 of the first $1{,}000$. 

The bit-stream can be compared to the binary expansion of prominent mathematical constants such as $\pi$, $e$, the golden ratio $\phi$, and $\sqrt{2}$. For reference, 

\texttt{ }

$\pi=$     \texttt{11.001001000011111101...}

$e=$       \texttt{10.101101111110000101...}

$\phi=$    \texttt{1.1001111000110111011...}

$1/\phi=$  \texttt{0.1001111000110111011...}

$\sqrt{2}=$\texttt{1.0110101000001001111...}

\texttt{ }

The first 500\,bits of the actual message are: 

\texttt{ }

\texttt{11101011\textcolor{red}{1}01\textcolor{red}{1}10101\textcolor{red}{0}111111110\textcolor{red}{1}\textcolor{red}{0}\textcolor{red}{1}011110000\textcolor{red}{0}0\textcolor{red}{1}\textcolor{red}{0}11}

\texttt{01001011\textcolor{red}{0}\textcolor{red}{1}00101001\textcolor{red}{0}11\textcolor{red}{1}011100\textcolor{red}{1}11110101100\textcolor{red}{1}101\textcolor{red}{1}}

\texttt{001110100\textcolor{red}{0}\textcolor{red}{0}101\textcolor{red}{1}00\textcolor{red}{0}0\textcolor{red}{1}1110000\textcolor{red}{1}10\textcolor{red}{0}011\textcolor{red}{1}0\textcolor{red}{0}10110100}

\texttt{\textcolor{red}{0}\textcolor{red}{1}01000101\textcolor{red}{0}110100101\textcolor{red}{0}000111\textcolor{red}{1}100011\textcolor{red}{1}\textcolor{red}{0}011\textcolor{red}{1}110\textcolor{red}{1}1}

\texttt{11000100010011\textcolor{red}{0}0011100\textcolor{red}{1}011\textcolor{red}{0}1\textcolor{red}{1}1\textcolor{red}{0}010\textcolor{red}{0}101000\textcolor{red}{1}010}

\texttt{0111000100101001010\textcolor{red}{0}001\textcolor{red}{1}1011\textcolor{red}{0}0\textcolor{red}{1}01010011010110}

\texttt{0\textcolor{red}{1}0110001\textcolor{red}{0}001101100100100\textcolor{red}{1}\textcolor{red}{1}10101\textcolor{red}{1}01100\textcolor{red}{1}01\textcolor{red}{0}100}

\texttt{110010\textcolor{red}{1}\textcolor{red}{1}00\textcolor{red}{1}00\textcolor{red}{1}0111\textcolor{red}{1}0\textcolor{red}{1}010\textcolor{red}{0}011111111010011\textcolor{red}{0}\textcolor{red}{1}100}

\texttt{0\textcolor{red}{1}01111010\textcolor{red}{1}0100100111001101101\textcolor{red}{0}10\textcolor{red}{1}1\textcolor{red}{0}10\textcolor{red}{0}0\textcolor{red}{0}1\textcolor{red}{1}00}

\texttt{0\textcolor{red}{0}\textcolor{red}{1}\textcolor{red}{0}\textcolor{red}{1}1\textcolor{red}{0}10\textcolor{red}{1}11001\textcolor{red}{1}00111\textcolor{red}{0}0\textcolor{red}{0}10\textcolor{red}{1}0\textcolor{red}{0}10011\textcolor{red}{1}0\textcolor{red}{0}111111\textcolor{red}{0}\textcolor{red}{1}}

\texttt{0\textcolor{red}{1}\textcolor{red}{0}001101\textcolor{red}{0}000000\textcolor{red}{0}\textcolor{red}{1}1010\textcolor{red}{1}10\textcolor{red}{0}1000101101010000011}

\texttt{11000}

\texttt{ }

As can be seen by comparing the prominent constants with the message, there is no straightforward match. Switching \texttt{0} and \texttt{1} does not improve the situation. I have also searched the On-Line Encyclopedia of Integer Sequences (OEIS\footnote{\url{https://oeis.org/}}) using various subsets of the message, but found nothing convincing. I have also repeated all tests after shifting the $\ell$ by 1.31 to approximate the infinite future, using linear interpolation.

As a further test for anomalies, I find that the distribution of binary digits is 254:246 in the first 500 values. This distribution passes the binomial fair coin tests. I find no significant autocorrelation. The series also passes the NIST battery of statistical tests for measuring the quality of a random number generator \citep{bassham2010sp}. This is an indication that there is no ``simple'' message in the number series, such as a subset that repeats (imperfectly due to measurement errors). These tests would, however, not find $\pi$ in the series, because $\pi$ is (conjectured to be) statistically random.

\section{Conclusion}
We may conclude that there is no obvious message on the CMB sky. Yet it remains unclear whether there is (was) a Creator, whether we live in a simulation, or whether the message is printed correctly in the previous section, but we fail to understand it.

\bibliography{references}

\begin{thebibliography}{}
\expandafter\ifx\csname natexlab\endcsname\relax\def\natexlab#1{#1}\fi
\providecommand{\url}[1]{\href{#1}{#1}}
\providecommand{\dodoi}[1]{}
\providecommand{\doeprint}[1]{}
\providecommand{\doarXiv}[1]{\href{https://arxiv.org/abs/#1}{\nolinkurl{https://arxiv.org/abs/#1}}}

\bibitem[{{Abitbol} {et~al.}(2020){Abitbol}, {Hill}, \&
  {Chluba}}]{2020ApJ...893...18A}
{Abitbol}, M.~H., {Hill}, J.~C., \& {Chluba}, J. 2020,
  \href{http://dx.doi.org/10.3847/1538-4357/ab7b70}{\color{magenta}\apj},
  \href{https://ui.adsabs.harvard.edu/abs/2020ApJ...893...18A}{\color{blue}893},
  \href{https://ui.adsabs.harvard.edu/abs/2020ApJ...893...18A}{\color{blue}18}

\bibitem[{{Adams} \& {Laughlin}(1997)}]{1997RvMP...69..337A}
{Adams}, F.~C., \& {Laughlin}, G. 1997,
  \href{http://dx.doi.org/10.1103/RevModPhys.69.337}{\color{magenta}Reviews of
  Modern Physics},
  \href{http://adsabs.harvard.edu/abs/1997RvMP...69..337A}{\color{blue}69},
  \href{http://adsabs.harvard.edu/abs/1997RvMP...69..337A}{\color{blue}337}

\bibitem[{{Bassham} {et~al.}(2010){Bassham}, {Rukhin}, {Soto}, {Nechvatal}, \&
  {Smid}}]{bassham2010sp}
{Bassham}, L.~E., {Rukhin}, A.~L., {Soto}, J., {et~al.} 2010, A statistical
  test suite for random and pseudorandom number generators for cryptographic
  applications (National Institute of Standards \& Technology)

\bibitem[{{Bennett} {et~al.}(2013){Bennett}, {Larson}, {Weiland}, {Jarosik},
  {Hinshaw}, {Odegard}, {Smith}, {Hill}, {Gold}, {Halpern}, {Komatsu}, {Nolta},
  {Page}, {Spergel}, {Wollack}, {Dunkley}, {Kogut}, {Limon}, {Meyer}, {Tucker},
  \& {Wright}}]{2013ApJS..208...20B}
{Bennett}, C.~L., {Larson}, D., {Weiland}, J.~L., {et~al.} 2013,
  \href{http://dx.doi.org/10.1088/0067-0049/208/2/20}{\color{magenta}\apjs},
  \href{https://ui.adsabs.harvard.edu/abs/2013ApJS..208...20B}{\color{blue}208},
  \href{https://ui.adsabs.harvard.edu/abs/2013ApJS..208...20B}{\color{blue}20}

\bibitem[{Bostrom(2003)}]{Bostrom2003}
Bostrom, N. 2003,
  \href{http://dx.doi.org/10.1111/1467-9213.00309}{\color{magenta}The
  Philosophical Quarterly}, 53, 243

\bibitem[{{Cruz} {et~al.}(2005){Cruz}, {Mart{\'\i}nez-Gonz{\'a}lez}, {Vielva},
  \& {Cay{\'o}n}}]{2005MNRAS.356...29C}
{Cruz}, M., {Mart{\'\i}nez-Gonz{\'a}lez}, E., {Vielva}, P., \& {Cay{\'o}n}, L.
  2005,
  \href{http://dx.doi.org/10.1111/j.1365-2966.2004.08419.x}{\color{magenta}\mnras},
  \href{https://ui.adsabs.harvard.edu/abs/2005MNRAS.356...29C}{\color{blue}356},
  \href{https://ui.adsabs.harvard.edu/abs/2005MNRAS.356...29C}{\color{blue}29}

\bibitem[{{Frolop} \& {Scott}(2016)}]{2016arXiv160309703F}
{Frolop}, A., \& {Scott}, D. 2016,
  \href{https://arxiv.org/abs/1603.09703}{\color{magenta}arXiv},
  \href{https://ui.adsabs.harvard.edu/abs/2016arXiv160309703F}{\color{blue}arXiv:1603.09703}.
\newblock \doarXiv{1603.09703}

\bibitem[{{Gurzadyan} \& {Penrose}(2010)}]{2010arXiv1011.3706G}
{Gurzadyan}, V.~G., \& {Penrose}, R. 2010,
  \href{https://arxiv.org/abs/1011.3706}{\color{magenta}arXiv},
  \href{https://ui.adsabs.harvard.edu/abs/2010arXiv1011.3706G}{\color{blue}arXiv:1011.3706}.
\newblock \doarXiv{1011.3706}

\bibitem[{{Hajian} \& {Souradeep}(2003)}]{2003ApJ...597L...5H}
{Hajian}, A., \& {Souradeep}, T. 2003,
  \href{http://dx.doi.org/10.1086/379757}{\color{magenta}\apjl},
  \href{https://ui.adsabs.harvard.edu/abs/2003ApJ...597L...5H}{\color{blue}597},
  \href{https://ui.adsabs.harvard.edu/abs/2003ApJ...597L...5H}{\color{blue}L5}

\bibitem[{{Hinshaw} {et~al.}(2013){Hinshaw}, {Larson}, {Komatsu}, {Spergel},
  {Bennett}, {Dunkley}, {Nolta}, {Halpern}, {Hill}, {Odegard}, {Page}, {Smith},
  {Weiland}, {Gold}, {Jarosik}, {Kogut}, {Limon}, {Meyer}, {Tucker}, {Wollack},
  \& {Wright}}]{2013ApJS..208...19H}
{Hinshaw}, G., {Larson}, D., {Komatsu}, E., {et~al.} 2013,
  \href{http://dx.doi.org/10.1088/0067-0049/208/2/19}{\color{magenta}\apjs},
  \href{https://ui.adsabs.harvard.edu/abs/2013ApJS..208...19H}{\color{blue}208},
  \href{https://ui.adsabs.harvard.edu/abs/2013ApJS..208...19H}{\color{blue}19}

\bibitem[{{Hsu} \& {Zee}(2006)}]{2006MPLA...21.1495H}
{Hsu}, S., \& {Zee}, A. 2006,
  \href{http://dx.doi.org/10.1142/S0217732306020834}{\color{magenta}Modern
  Physics Letters A},
  \href{http://adsabs.harvard.edu/abs/2006MPLA...21.1495H}{\color{blue}21},
  \href{http://adsabs.harvard.edu/abs/2006MPLA...21.1495H}{\color{blue}1495}

\bibitem[{{Jazayeri} {et~al.}(2017){Jazayeri}, {Sadr}, \&
  {Firouzjahi}}]{2017PhRvD..96b3512J}
{Jazayeri}, S., {Sadr}, A.~V., \& {Firouzjahi}, H. 2017,
  \href{http://dx.doi.org/10.1103/PhysRevD.96.023512}{\color{magenta}\prd},
  \href{https://ui.adsabs.harvard.edu/abs/2017PhRvD..96b3512J}{\color{blue}96},
  \href{https://ui.adsabs.harvard.edu/abs/2017PhRvD..96b3512J}{\color{blue}023512}

\bibitem[{{Krauss} \& {Scherrer}(2007)}]{2007GReGr..39.1545K}
{Krauss}, L.~M., \& {Scherrer}, R.~J. 2007,
  \href{http://dx.doi.org/10.1007/s10714-007-0472-9}{\color{magenta}General
  Relativity and Gravitation},
  \href{https://ui.adsabs.harvard.edu/abs/2007GReGr..39.1545K}{\color{blue}39},
  \href{https://ui.adsabs.harvard.edu/abs/2007GReGr..39.1545K}{\color{blue}1545}

\bibitem[{{Louis} {et~al.}(2019){Louis}, {Garrido}, {Soussana}, {Tristram},
  {Henrot-Versill{\'e}}, \& {Vanneste}}]{2019PhRvD.100b3518L}
{Louis}, T., {Garrido}, X., {Soussana}, A., {et~al.} 2019,
  \href{http://dx.doi.org/10.1103/PhysRevD.100.023518}{\color{magenta}\prd},
  \href{https://ui.adsabs.harvard.edu/abs/2019PhRvD.100b3518L}{\color{blue}100},
  \href{https://ui.adsabs.harvard.edu/abs/2019PhRvD.100b3518L}{\color{blue}023518}

\bibitem[{{Moss} {et~al.}(2008){Moss}, {Zibin}, \&
  {Scott}}]{2008PhRvD..77d3505M}
{Moss}, A., {Zibin}, J.~P., \& {Scott}, D. 2008,
  \href{http://dx.doi.org/10.1103/PhysRevD.77.043505}{\color{magenta}\prd},
  \href{https://ui.adsabs.harvard.edu/abs/2008PhRvD..77d3505M}{\color{blue}77},
  \href{https://ui.adsabs.harvard.edu/abs/2008PhRvD..77d3505M}{\color{blue}043505}

\bibitem[{{Mukherjee} \& {Souradeep}(2016)}]{2016PhRvL.116v1301M}
{Mukherjee}, S., \& {Souradeep}, T. 2016,
  \href{http://dx.doi.org/10.1103/PhysRevLett.116.221301}{\color{magenta}\prl},
  \href{https://ui.adsabs.harvard.edu/abs/2016PhRvL.116v1301M}{\color{blue}116},
  \href{https://ui.adsabs.harvard.edu/abs/2016PhRvL.116v1301M}{\color{blue}221301}

\bibitem[{{Planck Collaboration} {et~al.}(2018){Planck Collaboration},
  {Akrami}, {Arroja}, {Ashdown}, {Aumont}, {Baccigalupi}, {Ballardini},
  {Banday}, {Barreiro}, {Bartolo}, {Basak}, {Battye}, {Benabed}, {Bernard},
  {Bersanelli}, {Bielewicz}, {Bock}, {Bond}, {Borrill}, {Bouchet}, {Boulanger},
  {Bucher}, {Burigana}, {Butler}, {Calabrese}, {Cardoso}, {Carron},
  {Casaponsa}, {Challinor}, {Chiang}, {Colombo}, {Combet}, {Contreras},
  {Crill}, {Cuttaia}, {de Bernardis}, {de Zotti}, {Delabrouille}, {Delouis},
  {D{\'e}sert}, {Di Valentino}, {Dickinson}, {Diego}, {Donzelli}, {Dor{\'e}},
  {Douspis}, {Ducout}, {Dupac}, {Efstathiou}, {Elsner}, {En{\ss}lin},
  {Eriksen}, {Falgarone}, {Fantaye}, {Fergusson}, {Fernandez-Cobos}, {Finelli},
  {Forastieri}, {Frailis}, {Franceschi}, {Frolov}, {Galeotta}, {Galli},
  {Ganga}, {G{\'e}nova-Santos}, {Gerbino}, {Ghosh}, {Gonz{\'a}lez-Nuevo},
  {G{\'o}rski}, {Gratton}, {Gruppuso}, {Gudmundsson}, {Hamann}, {Hand ley},
  {Hansen}, {Helou}, {Herranz}, {Hivon}, {Huang}, {Jaffe}, {Jones}, {Karakci},
  {Keih{\"a}nen}, {Keskitalo}, {Kiiveri}, {Kim}, {Kisner}, {Knox},
  {Krachmalnicoff}, {Kunz}, {Kurki-Suonio}, {Lagache}, {Lamarre}, {Langer},
  {Lasenby}, {Lattanzi}, {Lawrence}, {Le Jeune}, {Leahy}, {Lesgourgues},
  {Levrier}, {Lewis}, {Liguori}, {Lilje}, {Lilley}, {Lindholm},
  {L{\'o}pez-Caniego}, {Lubin}, {Ma}, {Mac{\'\i}as-P{\'e}rez}, {Maggio},
  {Maino}, {Mand olesi}, {Mangilli}, {Marcos-Caballero}, {Maris}, {Martin},
  {Mart{\'\i}nez-Gonz{\'a}lez}, {Matarrese}, {Mauri}, {McEwen}, {Meerburg},
  {Meinhold}, {Melchiorri}, {Mennella}, {Migliaccio}, {Millea}, {Mitra},
  {Miville-Desch{\^e}nes}, {Molinari}, {Moneti}, {Montier}, {Morgante}, {Moss},
  {Mottet}, {M{\"u}nchmeyer}, {Natoli}, {N{\o}rgaard-Nielsen}, {Oxborrow},
  {Pagano}, {Paoletti}, {Partridge}, {Patanchon}, {Pearson}, {Peel}, {Peiris},
  {Perrotta}, {Pettorino}, {Piacentini}, {Polastri}, {Polenta}, {Puget},
  {Rachen}, {Reinecke}, {Remazeilles}, {Renzi}, {Rocha}, {Rosset}, {Roudier},
  {Rubi{\~n}o-Mart{\'\i}n}, {Ruiz-Granados}, {Salvati}, {Sandri}, {Savelainen},
  {Scott}, {Shellard}, {Shiraishi}, {Sirignano}, {Sirri}, {Spencer}, {Sunyaev},
  {Suur-Uski}, {Tauber}, {Tavagnacco}, {Tenti}, {Terenzi}, {Toffolatti},
  {Tomasi}, {Trombetti}, {Valiviita}, {Van Tent}, {Vibert}, {Vielva}, {Villa},
  {Vittorio}, {Wandelt}, {Wehus}, {White}, {White}, {Zacchei}, \&
  {Zonca}}]{2018arXiv180706205P}
{Planck Collaboration}, {Akrami}, Y., {Arroja}, F., {et~al.} 2018,
  \href{https://arxiv.org/abs/1807.06205}{\color{magenta}arXiv},
  \href{https://ui.adsabs.harvard.edu/abs/2018arXiv180706205P}{\color{blue}arXiv:1807.06205}.
\newblock \doarXiv{1807.06205}

\bibitem[{{Planck Collaboration} {et~al.}(2019{\natexlab{a}}){Planck
  Collaboration}, {Akrami}, {Ashdown}, {Aumont}, {Baccigalupi}, {Ballardini},
  {Band ay}, {Barreiro}, {Bartolo}, {Basak}, {Benabed}, {Bersanelli},
  {Bielewicz}, {Bock}, {Bond}, {Borrill}, {Bouchet}, {Boulanger}, {Bucher},
  {Burigana}, {Butler}, {Calabrese}, {Cardoso}, {Casaponsa}, {Chiang},
  {Colombo}, {Combet}, {Contreras}, {Crill}, {de Bernardis}, {de Zotti},
  {Delabrouille}, {Delouis}, {Di Valentino}, {Diego}, {Dor{\'e}}, {Douspis},
  {Ducout}, {Dupac}, {Efstathiou}, {Elsner}, {En{\ss}lin}, {Eriksen},
  {Fantaye}, {Fernandez-Cobos}, {Finelli}, {Frailis}, {Fraisse}, {Franceschi},
  {Frolov}, {Galeotta}, {Galli}, {Ganga}, {G{\'e}nova-Santos}, {Gerbino},
  {Ghosh}, {Gonz{\'a}lez-Nuevo}, {G{\'o}rski}, {Gruppuso}, {Gudmundsson},
  {Hamann}, {Hand ley}, {Hansen}, {Herranz}, {Hivon}, {Huang}, {Jaffe},
  {Jones}, {Keih{\"a}nen}, {Keskitalo}, {Kiiveri}, {Kim}, {Krachmalnicoff},
  {Kunz}, {Kurki-Suonio}, {Lagache}, {Lamarre}, {Lasenby}, {Lattanzi},
  {Lawrence}, {Le Jeune}, {Levrier}, {Liguori}, {Lilje}, {Lindholm},
  {L{\'o}pez-Caniego}, {Ma}, {Mac{\'\i}as-P{\'e}rez}, {Maggio}, {Maino}, {Mand
  olesi}, {Mangilli}, {Marcos-Caballero}, {Maris}, {Martin},
  {Mart{\'\i}nez-Gonz{\'a}lez}, {Matarrese}, {Mauri}, {McEwen}, {Meinhold},
  {Mennella}, {Migliaccio}, {Miville-Desch{\^e}nes}, {Molinari}, {Moneti},
  {Montier}, {Moss}, {Natoli}, {Pagano}, {Paoletti}, {Partridge}, {Perrotta},
  {Pettorino}, {Piacentini}, {Polenta}, {Puget}, {Rachen}, {Reinecke},
  {Remazeilles}, {Renzi}, {Rocha}, {Rosset}, {Roudier},
  {Rubi{\~n}o-Mart{\'\i}n}, {Ruiz-Granados}, {Salvati}, {Savelainen}, {Scott},
  {Shellard}, {Sirignano}, {Sunyaev}, {Suur-Uski}, {Tauber}, {Tavagnacco},
  {Tenti}, {Toffolatti}, {Tomasi}, {Trombetti}, {Valenziano}, {Valiviita}, {Van
  Tent}, {Vielva}, {Villa}, {Vittorio}, {Wandelt}, {Wehus}, {Zacchei}, {Zibin},
  \& {Zonca}}]{2019arXiv190602552P}
{Planck Collaboration}, {Akrami}, Y., {Ashdown}, M., {et~al.}
  2019{\natexlab{a}},
  \href{https://arxiv.org/abs/1906.02552}{\color{magenta}arXiv},
  \href{https://ui.adsabs.harvard.edu/abs/2019arXiv190602552P}{\color{blue}arXiv:1906.02552}.
\newblock \doarXiv{1906.02552}

\bibitem[{{Planck Collaboration} {et~al.}(2019{\natexlab{b}}){Planck
  Collaboration}, {Aghanim}, {Akrami}, {Ashdown}, {Aumont}, {Baccigalupi},
  {Ballardini}, {Banday}, {Barreiro}, {Bartolo}, {Basak}, {Benabed}, {Bernard},
  {Bersanelli}, {Bielewicz}, {Bock}, {Bond}, {Borrill}, {Bouchet}, {Boulanger},
  {Bucher}, {Burigana}, {Butler}, {Calabrese}, {Cardoso}, {Carron},
  {Casaponsa}, {Challinor}, {Chiang}, {Colombo}, {Combet}, {Crill}, {Cuttaia},
  {de Bernardis}, {de Rosa}, {de Zotti}, {Delabrouille}, {Delouis}, {Di
  Valentino}, {Diego}, {Dor{\'e}}, {Douspis}, {Ducout}, {Dupac}, {Dusini},
  {Efstathiou}, {Elsner}, {En{\ss}lin}, {Eriksen}, {Fantaye}, {Fernand
  ez-Cobos}, {Finelli}, {Frailis}, {Fraisse}, {Franceschi}, {Frolov},
  {Galeotta}, {Galli}, {Ganga}, {G{\'e}nova-Santos}, {Gerbino}, {Ghosh},
  {Giraud-H{\'e}raud}, {Gonz{\'a}lez-Nuevo}, {G{\'o}rski}, {Gratton},
  {Gruppuso}, {Gudmundsson}, {Hamann}, {Handley}, {Hansen}, {Herranz}, {Hivon},
  {Huang}, {Jaffe}, {Jones}, {Keih{\"a}nen}, {Keskitalo}, {Kiiveri}, {Kim},
  {Kisner}, {Krachmalnicoff}, {Kunz}, {Kurki-Suonio}, {Lagache}, {Lamarre},
  {Lasenby}, {Lattanzi}, {Lawrence}, {Le Jeune}, {Levrier}, {Lewis}, {Liguori},
  {Lilje}, {Lilley}, {Lindholm}, {L{\'o}pez-Caniego}, {Lubin}, {Ma},
  {Mac{\'\i}as-P{\'e}rez}, {Maggio}, {Maino}, {Mandolesi}, {Mangilli},
  {Marcos-Caballero}, {Maris}, {Martin}, {Mart{\'\i}nez-Gonz{\'a}lez},
  {Matarrese}, {Mauri}, {McEwen}, {Meinhold}, {Melchiorri}, {Mennella},
  {Migliaccio}, {Millea}, {Miville-Desch{\^e}nes}, {Molinari}, {Moneti},
  {Montier}, {Morgante}, {Moss}, {Natoli}, {N{\o}rgaard-Nielsen}, {Pagano},
  {Paoletti}, {Partridge}, {Patanchon}, {Peiris}, {Perrotta}, {Pettorino},
  {Piacentini}, {Polenta}, {Puget}, {Rachen}, {Reinecke}, {Remazeilles},
  {Renzi}, {Rocha}, {Rosset}, {Roudier}, {Rubi{\~n}o-Mart{\'\i}n},
  {Ruiz-Granados}, {Salvati}, {Sandri}, {Savelainen}, {Scott}, {Shellard},
  {Sirignano}, {Sirri}, {Spencer}, {Sunyaev}, {Suur-Uski}, {Tauber},
  {Tavagnacco}, {Tenti}, {Toffolatti}, {Tomasi}, {Trombetti}, {Valiviita}, {Van
  Tent}, {Vielva}, {Villa}, {Vittorio}, {Wandelt}, {Wehus}, {Zacchei}, \&
  {Zonca}}]{2019arXiv190712875P}
{Planck Collaboration}, {Aghanim}, N., {Akrami}, Y., {et~al.}
  2019{\natexlab{b}},
  \href{https://arxiv.org/abs/1907.12875}{\color{magenta}arXiv},
  \href{https://ui.adsabs.harvard.edu/abs/2019arXiv190712875P}{\color{blue}arXiv:1907.12875}.
\newblock \doarXiv{1907.12875}

\bibitem[{{Rath} {et~al.}(2018){Rath}, {Samal}, {Panda}, {Mishra}, \&
  {Aluri}}]{2018MNRAS.475.4357R}
{Rath}, P.~K., {Samal}, P.~K., {Panda}, S., {et~al.} 2018,
  \href{http://dx.doi.org/10.1093/mnras/sty007}{\color{magenta}\mnras},
  \href{https://ui.adsabs.harvard.edu/abs/2018MNRAS.475.4357R}{\color{blue}475},
  \href{https://ui.adsabs.harvard.edu/abs/2018MNRAS.475.4357R}{\color{blue}4357}

\bibitem[{{Sachs} \& {Wolfe}(1967)}]{1967ApJ...147...73S}
{Sachs}, R.~K., \& {Wolfe}, A.~M. 1967,
  \href{http://dx.doi.org/10.1086/148982}{\color{magenta}\apj},
  \href{https://ui.adsabs.harvard.edu/abs/1967ApJ...147...73S}{\color{blue}147},
  \href{https://ui.adsabs.harvard.edu/abs/1967ApJ...147...73S}{\color{blue}73}

\bibitem[{{Santos} {et~al.}(2012){Santos}, {Villela}, \&
  {Wuensche}}]{2012A&A...544A.121S}
{Santos}, L., {Villela}, T., \& {Wuensche}, C.~A. 2012,
  \href{http://dx.doi.org/10.1051/0004-6361/201118757}{\color{magenta}\aap},
  \href{https://ui.adsabs.harvard.edu/abs/2012A&A...544A.121S}{\color{blue}544},
  \href{https://ui.adsabs.harvard.edu/abs/2012A&A...544A.121S}{\color{blue}A121}

\bibitem[{{Schwarz} {et~al.}(2016){Schwarz}, {Copi}, {Huterer}, \&
  {Starkman}}]{2016CQGra..33r4001S}
{Schwarz}, D.~J., {Copi}, C.~J., {Huterer}, D., \& {Starkman}, G.~D. 2016,
  \href{http://dx.doi.org/10.1088/0264-9381/33/18/184001}{\color{magenta}Classical
  and Quantum Gravity},
  \href{https://ui.adsabs.harvard.edu/abs/2016CQGra..33r4001S}{\color{blue}33},
  \href{https://ui.adsabs.harvard.edu/abs/2016CQGra..33r4001S}{\color{blue}184001}

\bibitem[{{Scott} \& {Zibin}(2005)}]{2005physics..11135S}
{Scott}, D., \& {Zibin}, J.~P. 2005,
  \href{https://arxiv.org/abs/physics/0511135}{\color{magenta}arXiv},
  \href{https://ui.adsabs.harvard.edu/abs/2005physics..11135S}{\color{blue}physics/0511135}.
\newblock \doarXiv{physics/0511135}

\bibitem[{{Shaikh} {et~al.}(2019){Shaikh}, {Mukherjee}, {Das}, {Wand elt}, \&
  {Souradeep}}]{2019JCAP...08..007S}
{Shaikh}, S., {Mukherjee}, S., {Das}, S., {et~al.} 2019,
  \href{http://dx.doi.org/10.1088/1475-7516/2019/08/007}{\color{magenta}\jcap},
  \href{https://ui.adsabs.harvard.edu/abs/2019JCAP...08..007S}{\color{blue}2019},
  \href{https://ui.adsabs.harvard.edu/abs/2019JCAP...08..007S}{\color{blue}007}

\bibitem[{{Smoot} {et~al.}(1992){Smoot}, {Bennett}, {Kogut}, {Wright}, {Aymon},
  {Boggess}, {Cheng}, {de Amici}, {Gulkis}, {Hauser}, {Hinshaw}, {Jackson},
  {Janssen}, {Kaita}, {Kelsall}, {Keegstra}, {Lineweaver}, {Loewenstein},
  {Lubin}, {Mather}, {Meyer}, {Moseley}, {Murdock}, {Rokke}, {Silverberg},
  {Tenorio}, {Weiss}, \& {Wilkinson}}]{1992ApJ...396L...1S}
{Smoot}, G.~F., {Bennett}, C.~L., {Kogut}, A., {et~al.} 1992,
  \href{http://dx.doi.org/10.1086/186504}{\color{magenta}\apjl},
  \href{https://ui.adsabs.harvard.edu/abs/1992ApJ...396L...1S}{\color{blue}396},
  \href{https://ui.adsabs.harvard.edu/abs/1992ApJ...396L...1S}{\color{blue}L1}

\bibitem[{{Zibin} {et~al.}(2007){Zibin}, {Moss}, \&
  {Scott}}]{2007PhRvD..76l3010Z}
{Zibin}, J.~P., {Moss}, A., \& {Scott}, D. 2007,
  \href{http://dx.doi.org/10.1103/PhysRevD.76.123010}{\color{magenta}\prd},
  \href{https://ui.adsabs.harvard.edu/abs/2007PhRvD..76l3010Z}{\color{blue}76},
  \href{https://ui.adsabs.harvard.edu/abs/2007PhRvD..76l3010Z}{\color{blue}123010}

\bibitem[{{Zuntz} {et~al.}(2011){Zuntz}, {Zibin}, {Zunckel}, \&
  {Zwart}}]{2011arXiv1103.6262Z}
{Zuntz}, J., {Zibin}, J.~P., {Zunckel}, C., \& {Zwart}, J. 2011,
  \href{https://arxiv.org/abs/1103.6262}{\color{magenta}arXiv},
  \href{https://ui.adsabs.harvard.edu/abs/2011arXiv1103.6262Z}{\color{blue}arXiv:1103.6262}.
\newblock \doarXiv{1103.6262}

\end{thebibliography}
\end{document}